\documentclass[ACP]{copernicus}
\usepackage{amsmath}

\begin{document}
\newcommand{\beq}{\begin{equation}}
\newcommand{\eeq}{\end{equation}}
\newcommand{\pt}{\partial}

\title{Where do winds come from? A new theory on how water vapor
condensation influences atmospheric pressure and dynamics}

\author[1,2]{Anastassia M. Makarieva}
\author[1,2]{Victor G. Gorshkov}
\author[3,4]{Douglas Sheil}
\author[5,6]{Antonio D. Nobre}
\author[2]{Bai-Lian Li}

\affil[1]{Theoretical Physics Division, Petersburg Nuclear Physics Institute, 188300, Gatchina, St. Petersburg, Russia}
\affil[2]{XIEG-UCR International Center for Arid Land Ecology, University of California, Riverside, CA 92521, USA}
\affil[3]{Institute of Tropical Forest Conservation, Mbarara University of Science and Technology, Kabale, Uganda}
\affil[4]{Center for International Forestry Research, P.O. Box 0113 BOCBD, Bogor 16000, Indonesia}
\affil[5]{Centro de Ci\^{e}ncia do Sistema Terrestre INPE, S\~{a}o Jos\'{e} dos Campos SP 12227-010, Brazil}
\affil[6]{Instituto Nacional de Pesquisas da Amaz\^{o}nia, Manaus AM 69060-001, Brazil}


\runningtitle{Condensation-induced atmospheric dynamics}

\runningauthor{Makarieva et al.}

\correspondence{Anastassia M. Makarieva\\ (elba@peterlink.ru)}

\received{}
\pubdiscuss{} 
\revised{}
\accepted{}
\published{}


\firstpage{1}

\maketitle

\begin{abstract}
Phase transitions of atmospheric water play a ubiquitous role in the Earth's climate system, but their direct impact
on atmospheric dynamics has escaped wide attention. Here we examine and advance a theory as to how condensation influences atmospheric pressure
through the mass removal of water from the gas phase with a simultaneous account of the latent heat release.
Building from the fundamental physical principles we show that condensation is associated with a decline in air pressure in the lower atmosphere.
This decline occurs up to a certain height, which ranges from $3$ to $4$~km for
surface temperatures from 10 to 30~$^{\rm o}$C. We then estimate the horizontal pressure differences
associated with water vapor condensation and find that these are comparable in magnitude with
the pressure differences driving observed circulation patterns.
The water vapor delivered to the atmosphere via evaporation
represents a store of potential energy available to accelerate air and thus drive winds.
Our estimates suggest that the global mean power at which this potential energy is released by condensation
is around one per cent of the global solar power -- this is similar to
the known stationary dissipative power of general atmospheric circulation.
We conclude that condensation and evaporation merit attention as major, if previously overlooked,
factors in driving atmospheric dynamics.
\end{abstract}

\introduction

Phase transitions of water are among the major physical processes that shape
the Earth's climate. But such processes have not been well characterized.
This shortfall is recognized both as a challenge and a prospect for advancing our understanding of
atmospheric circulation  \citep[e.g.,][]{lor83,schn06}.
In {\it A History of Prevailing Ideas about the General Circulation of the Atmosphere} \citet{lor83}
wrote:

{\small "We may therefore pause and ask ourselves whether this step will be completed in the manner of the last three.
Will the next decade see new observational data that will disprove our present ideas? It would be difficult to show that this cannot happen.

Our current knowledge of the role of the various phases of water in the atmosphere is somewhat incomplete: eventually it must encompass
both thermodynamic and radiational effects. We do not fully understand the interconnections between the tropics, which contain the bulk of water,
and the remaining latitudes. . . . Perhaps near the end of the 20th century we shall suddenly discover that we are beginning the fifth step."}

\citet[Eq.~86]{lor67}, as well as several other authors after him \citep{tr87,tr91,gu91,ooy01,schu01,wack03,wack06}, recognized that local pressure is
reduced by precipitation and increased by evaporation.
\citet{qiu93} noted that "the mass depletion due to precipitation tends to reduce surface pressure,
which may in turn enhance the low-level moisture convergence and give a positive feedback to precipitation".
\citet{dool93} labeled the effect as a physically distinct "water vapor forcing". \citet{lack04} investigated the precipitation
mass sink for the case of Hurricane Lili (2002) and made an important observation that "the amount of atmospheric mass removed via precipitation
exceeded that needed to explain the model sea level pressure decrease".

Although the pressure changes associated with evaporation and condensation have received some attention, the investigations
have been limited: the effects remain poorly characterized in both theory and observations.
Previous investigations focused on temporal pressure changes not spatial gradients.
Even some very basic relationships remain subject to confusion.  For example, there is doubt as to whether condensation
leads to reduced or to {\it increased} atmospheric pressure \citep[p.~S12436]{poe09}.
Opining that the status of the issue in the meteorological literature is unclear,
\citet{hay09} suggested that to justify the claim of pressure reduction one would need to show that "the standard approaches
(e.g., set out in textbooks such as "Thermodynamics of Atmospheres and Oceans" by \citet{CW}) imply a drop in pressure associated with condensation".

Here we aim to clarify and describe, building from basic and
established physical principles, the pressure changes associated with
condensation. We will argue that atmospheric
water vapor represents a store of potential energy that becomes
available to accelerate air as the vapor condenses. Evaporation, driven by
the sun,
continuously replenishes the store of this energy in the atmosphere.

The paper is structured as follows. In Section~2 we analyze the process of adiabatic condensation
to show that it is always accompanied by a local decrease of air pressure. In Section~3 we evaluate the effects of water mass
removal and lapse rate change upon condensation in a vertical air column in approximate hydrostatic equilibrium. In Section~4 we
estimate the horizontal pressure gradients induced by water vapor condensation to show that these
are sufficient enough to drive the major circulation patterns on Earth (Section~4.1). We examine why the key relationships have remained unknown until
recently (Section~4.2).
We evaluate the mean global power available from condensation to drive the general atmospheric circulation (Secton~4.3).
Finally, we discuss the interplay between evaporation and condensation and the essentially different implications of their physics for
atmospheric dynamics (Section~4.4).
In the concluding section we discuss the importance of condensation as compared to differential heating as
the major driver of atmospheric circulation.
Our theoretical investigations strongly suggest that
the phase transitions of water vapor play a far greater role in driving atmospheric dynamics than is currently recognized.

\section{Condensation in a local air volume}

\subsection{Adiabatic condensation}

We will first show that adiabatic condensation is always accompanied by a decrease of air pressure in the local volume where it occurs.
The first law of thermodynamics for moist air saturated with water vapor reads \citep{gill,CW}
\beq
\label{1LT}
dQ = c_VdT + pdV + Ld\gamma ,
\eeq
\beq
\label{gamma}
\gamma \equiv \frac{p_v}{p} \ll 1,\,\,\, \frac{d\gamma}{\gamma} = \frac{dp_v}{p_v} - \frac{dp}{p}.
\eeq
Here $p_v$ is partial pressure of saturated water vapor, $p$ is air pressure, $T$ is absolute temperature,
$Q$ (J~mol$^{-1}$) is molar heat, $V$ (m$^3$~mol$^{-1}$) is molar volume, $L \approx 45$~kJ~mol$^{-1}$ is the molar heat of vaporization,
$c_V = \frac{5}{2}R$ is molar heat capacity of air at constant volume (J~mol$^{-1}$~K$^{-1}$), $R = 8.3$~J~mol$^{-1}$~K$^{-1}$ is the universal
gas constant. In processes not involving phase transitions the third term in (\ref{1LT}) is zero. In such processes partial pressure $p_v$ changes
proportionally to air pressure $p$, so that function $\gamma$ (\ref{gamma}) does not change.
The small value of $\gamma < 0.1$ under
terrestrial conditions allows us to
neglect the influence made by the heat capacity of liquid water in Eq.~(\ref{1LT}).

The partial pressure of saturated water vapor obeys the Clausius-Clapeyron equation:
\beq
\label{CC}
\frac{dp_v}{p_v} = \xi \frac{dT}{T},\,\,\, \xi \equiv \frac{L}{RT},
\eeq
\beq
\label{CCT}
p_v(T) = p_{v0}\exp (\xi_0 - \xi),
\eeq
where $p_{v0}$ and $\xi_0$ correspond to some reference temperature $T_0$. Below we use $T_0 = 303$~K and $p_{v0} = 42$~hPa
\citep{bo80} and neglect the dependence of $L$ on temperature.

We will also use the ideal gas law as the equation of state for atmospheric air:
\beq
\label{ig}
pV = RT,
\eeq
\beq
\label{Dig}
\frac{dp}{p} + \frac{dV}{V} = \frac{dT}{T}.
\eeq

Using Eq.~(\ref{Dig}) the first two terms in Eq.~(\ref{1LT}) can be written in the following form
\beq
\begin{split}
\label{cp}
c_VdT + pdV = \frac{RT}{\mu}\left( \frac{dT}{T} - \mu\frac{dp}{p} \right), \\
\mu \equiv \frac{R}{c_p} = \frac{2}{7} = 0.29, \,\,\, c_p = c_V + R.
\end{split}
\eeq

Writing $d\gamma$ in (\ref{1LT}) with use of (\ref{gamma}) and (\ref{CC}) as
\beq
\label{gamma2}
\frac{d\gamma}{\gamma} = \xi \frac{dT}{T} - \frac{dp}{p}
\eeq
and using the definition of $\xi$ (\ref{CC}) we arrive at the following form for the first law of thermodynamics
(\ref{1LT}):
\beq
\label{1LT2}
dQ = \frac{RT}{\mu} \left\{ \frac{dT}{T}(1 + \mu \gamma \xi^2) - \mu \frac{dp}{p} (1 + \gamma \xi) \right\}.
\eeq

In adiabatic processes $dQ = 0$, and the expression in braces in (\ref{1LT2}) turns to zero, which implies:
\beq
\label{fi}
\frac{dT}{T} = \frac{dp}{p} \varphi(\gamma, \xi), \,\,\, \varphi(\gamma,\xi) \equiv \mu \frac{1+ \gamma \xi}{1 + \mu \gamma \xi^2} \equiv \varphi.
\eeq
Note that $\mu$, $\gamma$ and $\xi$ are all dimensionless; $\gamma$ and $\xi$ are variables and
$\mu$ is a constant, $\varphi(0,0) = \mu$.
This is a general dependence of temperature on pressure in an adiabatic atmospheric process that involves
phase transitions of water vapor (evaporation or condensation), i.e. change of $\gamma$. At the same time
$\gamma$ itself is a function of temperature as determined by Eq.~(\ref{gamma2}):
\beq
\label{gamma3}
\frac{d\gamma}{\gamma} = \xi \frac{dT}{T} - \frac{dp}{p} = \frac{dT}{T} \frac{\xi\varphi - 1}{\varphi} = (\xi\varphi - 1) \frac{dp}{p}.
\eeq
One can see from Eqs.~(\ref{fi}) and (\ref{gamma3}) that the adiabatic phase transitions of water vapor are fully described
by the relative change of either pressure $dp/p$ or temperature $dT/T$. For the temperature range relevant for Earth
we have $\xi \equiv L/RT \approx 18$ so that
\beq
\label{ximu}
\xi \mu - 1 \approx 4.3.
\eeq

 Noting that $\mu$, $\gamma$, $\xi$ are all positive, from (\ref{fi}), (\ref{gamma3}) and (\ref{ximu}) we obtain
\beq
\label{proof}
\xi\varphi - 1 = \xi\mu \frac{1+\gamma\xi}{1 + \mu\gamma\xi^2} - 1 = \frac{\xi\mu - 1}{1+ \mu\gamma\xi^2} > 0.
\eeq

Condensation of water vapor corresponds to a decrease of $\gamma$, $d\gamma < 0$. It follows unambiguously
from Eqs.~(\ref{gamma3}) and (\ref{proof}) that if $d\gamma$ is negative, then $dp$ is negative too.
This proves that water vapor condensation in any adiabatic process
is necessarily accompanied by reduced air pressure.

\subsection{Adiabatic condensation cannot occur at constant volume}

Our previous result refutes the proposition that adiabatic condensation can lead to a pressure rise due to the release of latent heat
\citep[p.~S12436]{poe09}. Next, we show that while such a pressure rise was implied by calculations assuming {\it adiabatic condensation at constant volume},
in fact such a process is prohibited by the laws of thermodynamics and thus cannot occur.

Using (\ref{Dig}) and (\ref{fi}) we can express the relative change of molar volume $dV/V$ in terms of
$d\gamma/\gamma$:
\beq
\label{V}
\frac{dV}{V} = -\frac{1-\varphi}{\varphi\xi - 1} \frac{d\gamma}{\gamma}.
\eeq
Putting $dV = 0$ in (\ref{V}) we obtain
\beq
\label{0}
\frac{(1-\varphi)d\gamma}{(\xi\varphi - 1)\gamma} = 0.
\eeq
The denominator in (\ref{0}) is greater than zero, see Eq.~(\ref{proof}). In the numerator we note from
the definition of $\varphi$ (\ref{fi}) that $1 - \varphi = \frac{2\gamma}{7 +2\gamma\xi^2}\left[\frac{5}{2\gamma}+
\xi(\xi-1)\right]$. The expression in square brackets lacks real roots:
\beq
\frac{5}{2\gamma} + \xi^2 - \xi = 0, \,\,\, \xi = \frac{1}{2}\left(1\pm i\sqrt{\frac{10-\gamma}{\gamma}}\right),
\,\,\,\gamma \le 1.
\eeq
In consequence, Eq.~(\ref{0}) has a single solution $d\gamma = 0$. This proves that condensation cannot occur adiabatically at constant volume.

\subsection{Non-adiabatic condensation}

To conclude this section, we show that for any process where entropy increases, $dS = dQ/T >0$,
water vapor condensation ($d\gamma < 0$) is accompanied by drop of air pressure (i.e., $dp < 0$). We write the first law of thermodynamics
(\ref{1LT2}) and Eq.~(\ref{gamma3}) as
\beq
\label{S}
\frac{dS}{R} \frac{\mu}{1+\mu\gamma\xi^2} = \frac{dT}{T} - \varphi\frac{dp}{p},
\,\,\,\frac{dT}{T} = \frac{1}{\xi}\left(\frac{d\gamma}{\gamma}+\frac{dp}{p}\right).
\eeq
Excluding $dT/T$ from Eqs.~(\ref{S}) we obtain
\beq
\label{Sproof}
\frac{dp}{p} (\xi\varphi -1) = \frac{d\gamma}{\gamma} - \xi \frac{\mu}{1+\mu\gamma\xi}\frac{dS}{R}.
\eeq
The term in round brackets in Eq.~(\ref{Sproof}) is positive, see (\ref{proof}), the multiplier at $dS$ is also positive.
Therefore, when condensation occurs, i.e., when $d\gamma/\gamma < 0$, and $dS > 0$, the left-hand side of
Eq.~(\ref{Sproof}) is negative. This means that $dp/p < 0$, i.e., air pressure decreases.

Condensation can be accompanied by a pressure increase only if $dS < 0$. This requires that work is performed on the gas such as
occurs if it is isothermally compressed.
(We note too, that if pure saturated water vapor is isothermally compressed condensation occurs, but the Clausius-Clapeyron equation (\ref{CC}) shows that
the vapor pressure remains unchanged being purely a function of temperature.)

\section{Adiabatic condensation in the gravitational field}

\subsection{Difference in the effects of mass removal and temperature change on gas pressure in hydrostatic equilibrium}

We have shown that adiabatic condensation in any local volume is always accompanied by a drop of air pressure.
We will now explore the consequences of condensation for the vertical air column.

Most circulation patterns on Earth are much wider than high, with the ratio height/length being in the order of $10^{-2}$ for hurricanes
and down to $10^{-3}$ and below in larger regional circulations. As a consequence of mass balance, vertical velocity is smaller
than horizontal velocities by a similar ratio. Accordingly, the local pressure imbalances and resulting atmospheric accelerations are much smaller
in the vertical orientation than in the horizontal plane, the result being an atmosphere in approximate hydrostatic equilibrium \citep{gill}.
Air pressure then conforms to the equation
\beq
\label{HE}
-\frac{dp}{dz} = \rho g, \,\,\, p(0) \equiv p_s = g \int_0^\infty \rho(z)dz.
\eeq
Applying the ideal gas equation of state (\ref{ig}) we have from (\ref{HE})
\beq
\label{HEd}
\frac{dp}{dz} = -\frac{p}{h},\,\,\, h \equiv \frac{RT}{Mg}.
\eeq
This solves as
\beq
\label{HEz}
p(z) = p_s \exp\left\{-\int_0^z\frac{dz'}{h(z')} \right\}.
\eeq
Here $M$ is air molar mass (kg~mol$^{-1}$), which, as well as temperature $T(z)$, in the general case
also depends on $z$.

The value of $p_s$~(\ref{HE}), air pressure at the surface, appears as the constant of integration after Eq.~(\ref{HE}) is integrated over $z$.
It is equal to the weight of air molecules in the atmospheric column. It is important to bear in mind that $p_s$ does not depend on temperature,
but only on the amount of gas molecules in the column. It follows from this observation that any reduction of gas content in the column
reduces surface pressure.

Latent heat released when water condenses means that more energy has to be removed from a given volume of saturated air
for a similar decline in temperature when compared to dry air. This is why the moist adiabatic lapse rate is smaller
than the dry adiabatic lapse rate. Accordingly, given one and the same surface temperature $T_s$ in a column with rising air,
the temperature at some distance above the surface will be on average higher in a column of moist saturated air than in a dry one.

However, this does not mean that at a given height air pressure in the warmer column is greater than air pressure in the
colder column \citep[cf.][]{me09,mg09c}, because air pressure $p(z)$ (\ref{HEz}) depends on two parameters, temperature $T(z)$ and
surface air pressure (i.e., the total amount of air in the column). If the total amount of air in the warmer column is smaller
than in the colder column, air pressure in the surface layer will be lower in the warmer column despite its higher temperature.

In the following we estimate  the cumulative effect of
gas content and lapse rate changes upon condensation.

\subsection{Moist adiabatic temperature profile}

Relative water vapor content (\ref{gamma})
and temperature $T$ depend on height $z$. From Eqs.~(\ref{fi}),
(\ref{gamma3}) and (\ref{HEd}) we have
\beq
\label{Gm}
-\frac{dT}{dz} \equiv \Gamma = \varphi\frac{T}{h},\,\,\,\varphi \equiv \mu\frac{1+\gamma\xi}{1+\gamma\mu\xi^2},
\eeq
\beq
\label{gamma4}
-\frac{1}{\gamma}\frac{d\gamma}{dz} = \frac{\xi\varphi -1}{h}
\equiv\frac{\xi\mu-1}{1+\mu\gamma\xi^2}\frac{1}{h}.
\eeq
Eq.~(\ref{Gm}) represents the well-known formula for moist adiabatic gradient as given in \citet{glick}
for small $\gamma <0.1$. At $\gamma =0$ we have
$\varphi(\gamma,\xi) = \mu$ and $\Gamma_d = M_dg/c_p = 9.8$~K~km$^{-1}$, which is the dry adiabatic lapse rate
that is independent of height $z$, $M_d = 29$~g~mol$^{-1}$. For moist saturated air the change of temperature  $T$ and relative partial
pressure $\gamma$ of water vapor with height is determined by the system of differential equations (\ref{Gm}), ({\ref{gamma4}).

Differentiating both parts of Clapeyron-Clausius equation (\ref{CC}) over $z$ we have, see~(\ref{Gm}):
\beq
\begin{split}
\label{CCz}
\frac{dp_v}{dz} = -\frac{p_v}{h_v},\,\,\, h_v \equiv \frac{RT^2}{L\Gamma} = \frac{T}{\xi\Gamma} = \frac{h}{\xi\varphi},\\
p_v(z) = p_{vs}\exp\left\{-\int_0^z\frac{dz'}{h_v}\right\} , \,\,\, p_{vs} \equiv p_v(0).
\end{split}
\eeq
The value of $h_v$ represents a fundamental scale height for the vertical distribution of saturated water vapor.
At $T_s = 300$~K this height $h_v$ is approximately $4.5$~km.

Differentiating both parts of Eq.~(\ref{gamma}) over $z$ with use of (\ref{HEd}) and (\ref{CCz})
and noticing that $h_v = h/(\xi \varphi)$ we have
\beq
\label{h}
-\frac{1}{\gamma} \frac{d\gamma}{dz} = \frac{1}{p_v}\frac{dp_v}{dz}-\frac{1}{p}\frac{dp}{p} = \frac{1}{h_v}-\frac{1}{h} \equiv \frac{1}{h_\gamma},\,
h_\gamma \equiv \frac{h_vh}{h-h_v}.
\eeq
This equation is equivalent to Eq.~(\ref{gamma4}) when Eqs.~(\ref{Gm}) and (\ref{CCz}) are taken into account.
Height $h_\gamma$ represents the vertical scale of the condensation process.
Height scales $h_v$ (\ref{CCz}) and $h_\gamma$ (\ref{h}) depend on $\varphi(\gamma,\xi)$ (\ref{Gm}) and, consequently, on
$\gamma$. At $T_s = 300$~K height $h_\gamma \approx 9$~km, in close proximity to the water vapor scale height described
by \citet{map01}.

\subsection{Pressure profiles in moist versus dry air columns}

We start by considering two static vertically isothermal atmospheric columns of unit area, A and B, with temperature $T(z) = T_s$ independent of height.
Column A contains moist air with water vapor saturated at the surface, column B contains dry air only. Surface temperatures and surface
pressures in the two columns are equal.
In static air Eq.~(\ref{HE}) is exact and applies to each component of the gas mixture as well as to the mixture as a whole.
At equal surface pressures, the total air mass and air weight are therefore the same in both columns.
Water vapor in column A is saturated at the surface (i.e., at $z = 0$) but non-saturated above it (at $z > 0$).
The saturated partial pressure of water vapor at the surface $p_{v}(T_s)$ (\ref{CCT}) is determined by surface temperature
and, as it is in hydrostatic equilibrium, equals the weight of water vapor in the static column.

We now introduce a non-zero lapse rate to both columns: the moist adiabatic $\Gamma$ (\ref{Gm}) to column A
and the dry adiabatic $\Gamma_d$ in column B. (Now the columns cannot be static: the adiabatic lapse rates
are maintained by the adiabatically ascending air.) Due to the decrease of temperature with height,
some water vapor in column A undergoes condensation. Water vapor becomes saturated everywhere in the column (i.e., at $z \ge 0$), with pressure
$p_v(z)$ following Eq.~(\ref{CCz}) and density $\rho_v = p_vM_v/(RT) = p_v/(gh_n)$ following
\beq
\label{rho_v}
\begin{split}
\rho_v(z) = \rho_v(T_s)\frac{h_{ns}}{h_n(z)}\exp\left\{-\int_0^z\frac{dz'}{h_v(z')}\right\},\\
\rho_v(T_s) \equiv \frac{p_v(T_s)}{gh_n(T_s)},\,\,h_n \equiv \frac{RT(z)}{M_vg}, \,\,T(z) = T_s - \Gamma z.
\end{split}
\eeq
Here $h_n(z)$ is the scale height of the hydrostatic distribution of water vapor
in the isothermal atmosphere with $T_s = T(z)$.

\begin{figure*}[t]
\label{DpF}
\vspace*{2mm}
\begin{center}
\includegraphics[width=0.9\textwidth]{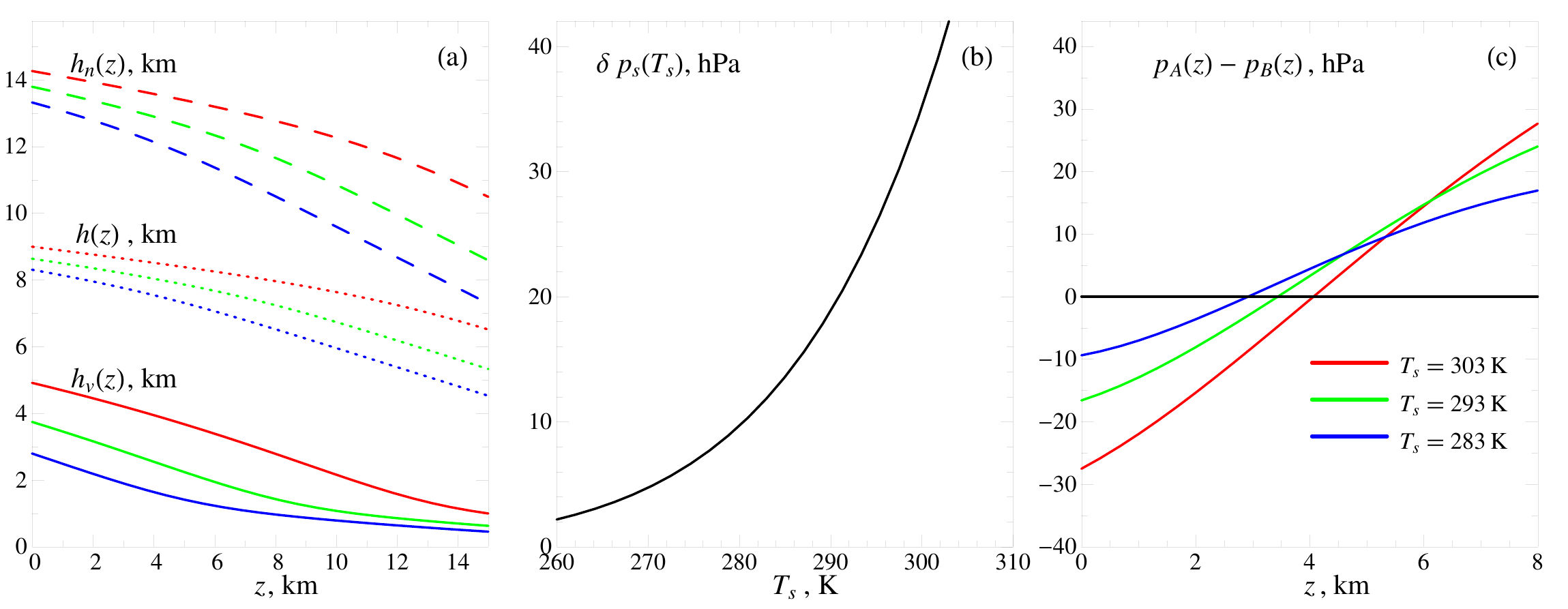}
\end{center}
\caption{(a): scale height of saturated water vapor $h_v(z)$ (\ref{CCz}), hydrostatic scale height of water vapor $h_n(z)$ (\ref{rho_v}),
and scale height of moist air $h(z)$ (\ref{HEd}) in the column with moist adiabatic lapse rate (\ref{Gm}) for three values of surface temperature $T_s$;
(b): condensation-induced drop of air pressure at the surface (\ref{Dp}) as dependent on surface temperature $T_s$; (c):
pressure difference versus altitude $z$ between atmospheric columns A and B with moist and dry adiabatic lapse rates, Eqs.~(\ref{resA}),
(\ref{resB}), respectively, for three values of surface temperature $T_s$. Height $z_c$ at which $p_A(z_c) - p_B(z_c) = 0$ is 2.9, 3.4 and 4.1~km
for $283$, $293$ and $303$~K, respectively.
Due to condensation, at altitudes below $z_c$ the air pressure is lower in column A despite it being warmer than column B.}
\end{figure*}

The change in pressure $\delta p_s$ in column~A due to water vapor condensation
is equal to the difference between the initial weight of water vapor $p_v(T_s)$
and the weight of saturated water vapor:
\beq
\label{Dp}
\begin{split}
\delta p_s = p_v(T_s) - g \int_0^\infty\rho_v(z)dz \le p_v(T_s) - \rho_v(T_s) g h_v(T_s) = \\
= p_v(T_s) \left(1 - \frac{h_{vs}}{h_{ns}}\right) = p_v(T_s)\left(1 - \frac{M_vgT_s}{L\Gamma_s} \right).
\end{split}
\eeq
\noindent
The inequality in Eq.~(\ref{Dp}) represents a conservative estimate of $\delta p_s$ due to the approximation $h_v(z) = h_v(T_s)$ made
while integrating $\rho_v(z)$ (\ref{rho_v}). As far as $h_v(z)$ declines with height
more rapidly than $h_n(z)$, Fig.~1a, the exact magnitude of this integral
is smaller, while the value of $\delta p_s$ is larger. The physical meaning of estimate (\ref{Dp}) consists in the fact
that the drop of temperature with height compresses the water vapor distribution $h_{ns}/h_{vs}$-fold compared to the
hydrostatic distribution \citep{mg07,mg09a}.

The value of $\delta p_s$ (\ref{Dp}) was calculated as the difference between
the weight per unit surface area of vapor in the isothermal hydrostatic column and the weight of water vapor that
condensed when a moist adiabatic lapse rate was applied. This derivation can also be understood in terms of the variable
conventionally called the {\it adiabatic liquid water content} \citep[e.g.,][Eq.~6.41]{CW}. We can represent the total mixing
ratio of moisture (by mass) as $q_t \equiv q_v + q_l = \rho_v/\rho + \rho_l /\rho$,
where $\rho_v$ is the mass of vapor $q_t \ll 1$ and $\rho_l$ is the mass of liquid water per unit air volume; $q_t \ll 1$.
The total adiabatic liquid water content in the column equals the integral of $q_l\rho$ over $z$
at constant $q_t$, $q_l \rho = q_t \rho - q_v \rho = q_t \rho - \rho_v$. The value of $\delta p_s$ (\ref{Dp}) is equal
to this integral (mass per unit area) multiplied by the gravitational acceleration (giving weight per unit area):
\beq
\label{rho}
\delta p_s = g \int_0^\infty q_l \rho dz = g \left( \int_0^\infty q_t \rho dz - \int_0^\infty \rho_v dz \right).
\eeq
The first integral in the right-hand part of this equation gives the mass of vapor in the considered atmospheric
column if water vapor were a non-condensable gas, $q_v = q_t = \rm {const}$. This term is analagous to the first term, $p_v(T_s)$, in the right-hand
side of Eq.~(\ref{Dp}), where a static isothermal column was considered.
The second term is identical to the second term, $g\int_0^\infty \rho_v dz$, in Eq.~(\ref{Dp}).

Using the definition of $h_v(T_s)$~(\ref{CCz}), $h_n(T_s)$~(\ref{rho_v}) and recalling that $M_v/M_d = 0.62$ and
$p_v(T_s) = \gamma_s p_s$, see~(\ref{CCT}), we obtain the following expression for the $\delta p_s$ estimate~(\ref{Dp}), Fig.~1b:
\beq
\label{Dp2}
\frac{\delta p_s}{p_s} \approx \gamma_s \left( 1 - 0.62\frac{1+\gamma_s\mu\xi_s^2}{\mu\xi_s + \gamma_s \mu \xi_s^2} \right).
\eeq
\noindent
Note that $\delta p_s/p_s$ is proportional to $\gamma_s$ and increases exponentially with the rise of temperature.

After an approximate hydrostatic equilibrium is established, the vertical pressure profiles for columns~A and B become, cf.~({\ref{HEz}):
\beq
\label{resA}
p_A(z) = p_s (1 - \frac{\delta p_s}{p_s}) \exp\left\{-\int_0^z\frac{dz'}{h_A(z')}\right\}, \,\,\,h_A \equiv \frac{RT}{Mg};
\eeq
\beq
\label{resB}
p_B(z) = p_s \exp\left\{-\int_0^z\frac{dz'}{h_B(z')}\right\}, \,\,\,h_B \equiv \frac{RT_d}{M_dg}.
\eeq
Here $M(z) = M_d (1 - \gamma) + M_v\gamma$; $\gamma \equiv p_v(z)/p_A(z)$ and $T(z)$ obey Eqs.~(\ref{Gm}) and
(\ref{gamma4}), $T_d(z) \equiv T_s - \Gamma_d z$.

In Fig.~1c the difference $p_A(z)- p_B(z)$ is plotted for three surface temperatures, $T_s = 10^{\rm o}$,
$20^{\rm o}$ and $30^{\rm o}$~C. In all three cases condensation has resulted in a lower air pressure in column~A
compared to column~B everywhere below $z_c \approx 2.9$, $3.4$ and $4.1$~km, respectively. It is only above that height
that the difference in lapse rates makes pressure in the moist column higher than in the dry column.

\section{Relevance of the condensation-induced pressure changes for atmospheric processes}

\subsection{Horizontal pressure gradients associated with vapor condensation}

We have shown that condensation of water vapor produces a drop of air pressure in the lower atmosphere up to an altitude of a few kilometers, Fig.~1c,
in a moist saturated hydrostatically adjusted column.
In the dynamic atmospheric context the vapor condenses and latent heat is released during the ascent of moist air.
The vertical displacement of air is inevitably accompanied by its horizontal displacement. This translates much of the condensation-induced pressure
difference to a horizontal pressure gradient.
Indeed, as the upwelling air loses its water vapor, the surface pressure diminishes via hydrostatic adjustment producing
a surface gradient of total air pressure between the areas of ascent and descent. The resulting horizontal pressure gradient is proportional to the
the ratio of vertical to horizontal velocity $w/u$ \citep{mg09b}.

We will illustrate this point regarding the magnitude of the resulting atmospheric pressure gradient
for the case of a stationary axis-symmetric circulation developing above a horizontally isothermal oceanic
surface. In cylindrical coordinates the
continuity equation for the mixture of condensable (vapor) and non-condensable (dry air) gases can be written as
\beq
\label{Nd}
\frac{1}{r}\frac{\pt (N_d u r)}{\pt r} + \frac{\pt (N_d w)}{\pt z} = 0;
\eeq
\beq
\label{Nv}
\frac{1}{r}\frac{\pt (N_v u r)}{\pt r} + \frac{\pt (N_v w)}{\pt z} = S(r,z);
\eeq
\beq
\label{src}
S(r,z) = w\left(\frac{\pt N_v}{\pt z} - \frac{N_v}{N}\frac{\pt N}{\pt z}\right) = w\gamma\frac{\pt \gamma}{\pt z},\,\,\,N = N_v + N_d.
\eeq
Here $N_d$ and $N_v$ are molar densities of dry air and water vapor, respectively; $\gamma \equiv N_v/N$, see~(\ref{gamma}),
$r$ is the distance from the center of the area
where condensation takes place, $S(r,z)$ is the sink term describing the non-conservation of the condensable
component (water vapor). Saturated pressure of water vapor depends on temperature alone. Assuming that vapor is saturated at the isothermal surface
we have $\pt N_v/\pt r = 0$, so $N_v$ only depends on $z$. (Note that this condition necessitates either that there is
an influx of water vapor via evaporation from the surface (if the circulation pattern is immobile), or that
the pressure field moves as vapor is locally depleted.
The second case occurs in compact circulation patterns like hurricanes and tornadoes\footnote{\label{arx}Makarieva, A. M. and Gorshkov, V. G.: Potential energy of atmospheric water vapor
and the air motions induced by water vapor condensation on different spatial scales, http://arxiv.org/abs/1003.5466 [physics.gen-ph], 2010.}.)
As the air ascends with vertical velocity $w$, vapor molar density decreases due to condensation
and due to the expansion of the gas along the vertical gradient of decreasing pressure. The latter effect equally influences all gases, both condensable
and non-condensable. Therefore, the volume-specific rate $S(r,z)$ at which vapor molecules are locally removed from the gaseous phase is equal to
$w(\pt N_v / \pt z - (N_v / N ) \pt N / \pt z)$, see~(\ref{1LT}), (\ref{gamma}). The second term describes the expansion of vapor at a constant mixing ratio
which would have occurred if vapor were non-condensable as the other gases. (If vapor did not condense, its density would decrease
with height as a constant proportion of the total molar density of moist air as with any other atmospheric gas.)

The mass of dry air is conserved, Eq.~(\ref{Nd}). Using this fact, Eq.~(\ref{src}) and $\pt N_v/\pt r = 0$ one can see
that
\beq
\label{int}
N \left(\frac{1}{r}\frac{\pt (ur)}{\pt r} + \frac{\pt w}{\pt z}\right) + w\frac{\pt N}{\pt z} = 0.
\eeq
Now expressing $\pt N/\pt r = \pt N_d/\pt r + \pt N_v/\pt r$ from Eqs.~(\ref{Nd}) and (\ref{Nv}) with use of Eq.~(\ref{int}) we obtain
\beq
\label{gr1}
\frac{\pt N}{\pt r} = \frac{w}{u}\left(\frac{\pt N_v}{\pt z} - \frac{N_v}{N} \frac{\pt N}{\pt z}\right).
\eeq
Using the equation of state for moist air $p = NRT$ and water vapor $p_v = N_vRT$
we obtain from Eqs.~(\ref{gr1}) and (\ref{h}):
\beq
\label{gr2}
\frac{\pt p}{\pt r} = \left(\frac{\pt p_v}{\pt z} - \frac{p_v}{p} \frac{\pt p}{\pt z}\right)\frac{w}{u} =
\frac{\gamma p}{h_\gamma}\frac{w}{u}.
\eeq
Here velocities $w$ and $u$ represent vertical and radial velocities of the ascending air flow, respectively.
The ascending air converges towards the center of the area where condensation occurs. Scale height $h_\gamma$ is defined in Eq.~(\ref{h}).
A closely related formula for horizontal pressure gradient can be applied to a linear two-dimensional air flow,
with $\pt p/\pt r$ replaced by $\pt p/\pt x$.

Equation~(\ref{gr2}) shows that the difference between the scale heights $h_v$ and $h$ (\ref{h})
of the {\it vertical} pressure distributions for water vapor and moist air leads to the appearance of
a {\it horizontal} pressure gradient of moist air as a whole. This equation contains the ratio of vertical
to horizontal velocity. Estimating this ratio it is possible to evaluate, for a given circulation,
what sorts of horizontal pressure gradients are produced by condensation and whether these gradients
are large enough to maintain the observed velocities via the positive physical feedback described by Eq.~(\ref{gr2}).

For example, for Hadley cells at $T = 300$~K, $h_\gamma = 9$~km, $\gamma = 0.04$ and
a typical ratio of $w/u \sim 10^{-3}$ we obtain from Eq.~(\ref{gr2}) a pressure gradient of about $0.4$~Pa~km$^{-1}$.
On a distance of 3000 km such a gradient would correspond to a pressure difference of 12 hPa,
which is close to the upper range of the actually observed pressure differences in the region
\citep[e.g.,][Fig.~1]{mu88}.
This estimate illustrates our proposal that condensation should be considered one of the main determinants of
atmospheric pressure gradients and, hence, air circulation.

Similar pressure differences and gradients, also comparable in magnitude to
$\delta p_s$ (\ref{Dp}) and $\pt p /\pt r$ (\ref{gr2}) are observed within cyclones, both tropical and extratropical,
and persistent atmospheric patterns in the low latitudes \citep{ho80,zhou98,bru00,nic00,sim08}.
For example, the mean depth of Arctic cyclones, 5 hPa \citep{sim08}, is about ten times smaller than the mean depth
of a typical tropical cyclone \citep{ho80}. This pattern agrees well with the
Clausius-Clapeyron dependence of $\delta p_s$, Fig.~1b, which would predict an 8 to 16-fold decrease with mean oceanic temperature
dropping by 30-40 degrees Celsius. The exact magnitude of pressure gradient and the resulting velocities will depend
on the horizontal size of the circulation pattern, the magnitude of friction
and degree of the radial symmetry \citep{mg09a,mg09b}$^{\ref{arx}}$.

\subsection{Regarding previous oversight of the effect}

For many readers a major barrier to acceptance of our propositions may be to understand how such a fundamental
physical mechanism has been overlooked until now. Why has this theory come to light only now in what is widely regarded as a mature field?
We can offer a few thoughts based on our readings and discussions with colleagues.

The condensation-induced pressure gradients that we have been examining are associated with density
gradients that have been conventionally considered as minor and thus ignored in the continuity equation \citep[e.g.,][]{sa08}.
For example, a typical $\Delta p = 50$~hPa pressure difference observed along the horizontally isothermal
surface between the outer environment and the hurricane center \citep[e.g.,][]{ho80}
is associated with a density difference of only around 5\%.
This density difference can be safely neglected when estimating the resulting air velocity $u$ from
the known pressure differences $\Delta p$. Here the basic scale relation is given by
Bernoulli's equation, $\rho u^2/2 = \Delta p$. The point is that a 5\% change in $\rho$ does not significantly impact the magnitude of the estimated air velocity {\it at a given} $\Delta p$.
But, as we have shown in the previous section, for the determination of the pressure gradient (\ref{gr2}) the density
difference and gradient (\ref{gr1}) are key.

Considering the equation of state (\ref{ig}) for the horizontally isothermal surface
we have $p=C\rho$, where $C \equiv RT/M = {\rm const.}$
Irrespective of why the considered pressure difference arises, from Bernoulli's equation we know that
$u^2 = 2\Delta p/\rho = 2C \Delta \rho/\rho$, $\Delta \rho = \rho_0 - \rho$.
Thus, if one puts $\Delta \rho/\rho = \Delta p/p$ equal to zero, no velocity forms and there is no circulation.
Indeed, we have $u^2 = 2\Delta p/\rho =
2C\Delta \rho/\rho = 2C (\Delta \rho/\rho_0) (1+\Delta \rho/\rho_0 + ...)$. As one can see,
discarding $\Delta \rho$ compared to $\rho$ does indeed correspond to discarding the
higher order term of the smallness parameter $\Delta \rho/\rho$.
But with respect to the pressure gradient, the main effect is proportional to the smallness parameter $\Delta \rho/\rho_0$
itself. If the latter is assumed to be zero, the effect is overlooked.
We suggest that this dual aspect of the magnitude of condensation-related density changes has not been recognized and this has contributed
to the neglect of condensation-associated pressure gradients in the Earth's atmosphere.

Furthermore, the consideration of air flow associated with phase transitions
of water vapor has been conventionally reduced to the consideration of the net fluxes of matter. Suppose we have a linear circulation pattern
divided into the ascending and descending parts, with similar evaporation rates $E$ (kg~H$_2$O~m$^{-2}$~s$^{-1}$) in both regions.
In the region of ascent the water vapor precipitates at a rate $P$. This creates a mass sink $E - P$, which has to be balanced by water vapor
import from the region of descent. Approximating the two regions as boxes of height $h$, length $l$ and width $d$,
the horizontal velocity $u_t$ associated with this mass transport can be estimated from the mass balance equation
\beq
\label{mb}
(P - E)ld = u_t \rho h d,\,\,\,u_t = \frac{(P - E)}{\rho}\frac{l}{h}.
\eeq
\noindent
Equation~(\ref{mb}) says that the depletion of air mass in the region of ascent at a total rate of $(P-E)ld$ is
compensated for by the horizontal air influx from the region of descent that goes with velocity $u_t$
via vertical cross-section of area $hd$.
For typical values in the tropics with $P - E \sim 5$~mm~day$^{-1} = 5.8 \times 10^{-5}$~kg~H$_2$O~m$^{-2}$~s$^{-1}$ and $l/h \sim 2 \times 10^3$
we obtain $u_t \sim 1$~cm~s$^{-1}$. For regions where precipitation and evaporation are smaller, the
value of $u_t$ will be smaller too. For example, \citet[p.~51]{lor67} estimated $u_t$ to be $\sim 0.3$~cm~s$^{-1}$
for the air flow across latitude 40$^{\rm o}$S.

With $\rho \approx \rho_d$ the value of $u_t$ can be understood as the
mass-weighted horizontal velocity of the dry air + water vapor mixture, which is
the so-called barycentric velocity, see, e.g., \citep{wack03,wack06}. There is no net flux of dry air between the regions
of ascent and descent, but there is a net flux of water vapor from the region of descent to the region of ascent. This
leads to the appearance of a non-zero horizontal velocity $u_t$ directed towards the region of ascent.
Similarly, vertical barycentric velocity at the surface is $w_t \approx (E-P)/\rho$ \citep{wack03},
which reflects the fact that there is no net flux of dry air via the Earth's surface, while water vapor is added
via evaporation or removed through precipitation. The absolute magnitude of vertical barycentric velocity $w_t$
for the calculated tropical means is vanishingly small, $w_t \sim 0.05$~mm~s$^{-1}$.

We speculate that the low magnitude of barycentric velocities has contributed to the
judgement that water's phase transitions cannot be a major driver of atmospheric {\it dynamics}.
However, barycentric velocities should not be confused \citep[e.g.,][]{me09} with the actual air velocities.
Unlike the former, the latter cannot be estimated without considering atmospheric pressure
gradients \citep{mg09c}. For example, in the absence of friction, the maximum linear velocity $u_c$ that could
be produced by condensation in a linear circulation pattern in the tropics constitutes
\beq
\label{uc}
u_c = \sqrt{2\Delta p/\rho} \sim 45{\rm\,m\,s^{-1}} \gg u_t.
\eeq
\noindent
Here $\Delta p$ was taken equal to 12~hPa as estimated from Eq.~(\ref{gr2}) for Hadley cell in Section~4.1. As one can see, $u_c$
(\ref{uc}) is much greater than $u_t$~(\ref{mb}). As some part of potential energy associated with the condensation-induced
pressure gradient is lost to friction \citep{mg09a}, real air velocities observed in large-scale circulation are an order of magnitude smaller
than $u_c$, but still nearly three orders of magnitude greater than $u_t$.

\subsection{The dynamic efficiency of the atmosphere}

We will now present another line of evidence for the importance of condensation-induced dynamics:  we shall show that it offers an improved
 understanding of the efficiency with which  the Earth's
atmosphere can convert solar energy into kinetic energy of air circulation.
While the Earth on average absorbs
about $I \approx 2.4\times 10^2$~W~m$^{-2}$ of solar radiation \citep{rav89}, only a minor part $\eta \sim 10^{-2}$ of this energy is
converted to the kinetic power of atmospheric and oceanic movement. \citet[p.~97]{lor67} notes, "the determination and explanation of
efficiency $\eta$ constitute the fundamental observational and theoretical problems of atmospheric energetics".
Here the condensation-induced dynamics yields a relationship that is quantitative in nature and can be estimated
directly from fundamental atmospheric parameters.

A pressure gradient is associated with a store of potential energy. The physical dimension of pressure gradient
coincides with the dimension of force per unit air volume, i.e. $1$~Pa~m$^{-1}$ = $1$~N~m$^{-3}$. When an air parcel
moves along the pressure gradient, the potential energy of the pressure field is converted to the kinetic energy.
The dimension of pressure is identical to the dimension of energy density:
1~Pa = 1~N~m$^{-2}$ = 1~J~m$^{-3}$. As the moist air in the lower
part of the atmospheric column rises to height $h_\gamma$ where most part of its water vapor condenses, the potential energy released
amounts to approximately $\delta p_s$~(\ref{Dp}). The potential energy released $\pi_v$ per unit mass of water vapor condensed,
dimension J~(kg~H$_2$O~)$^{-1}$, thus becomes
\beq
\label{pi}
\pi_v(T_s) = \frac{\delta p_s}{\rho_v} = \frac{RT_s}{M_v}\left(1 - \frac{M_vgT_s}{L\Gamma_s}\right).
\eeq
The global mean precipitation rate is $P \sim 10^3$~kg~H$_2$O~m$^{-2}$~year$^{-1}$ \citep{lv},
global mean surface temperature is $T_s = 288$~K and the observed mean tropospheric lapse rate $\Gamma_o = 6.5$~K~km$^{-1}$ \citep{glick}.
Using these values and putting $\Gamma_o$ instead of the moist adiabatic lapse rate $\Gamma_s$ in~(\ref{pi}),
we can estimate the global mean rate $\Pi_v = P\pi_v$ at which the condensation-related potential energy is available for conversion into kinetic energy.
At the same time we also estimate the efficiency $\eta = \Pi_v/I$ of atmospheric circulation that can be generated by solar energy via the condensation-induced
pressure gradients:
\beq
\label{Pi}
\Pi_v = P \pi_v \sim 3.5~{\rm W~m^{-2}},\,\,\,\eta \sim 0.015.
\eeq
Thus, the proposed approach not only clarifies the dynamics of solar energy conversion to the kinetic power of air movement (solar power spent on
evaporation $\to$ condensation-related release of potential power $\to$ kinetic power generation). It does so in a quantiatively tractable
manner explaining the magnitude of the dissipative power associated with maintaining the kinetic energy of
the Earth's atmosphere.

Our estimate of atmospheric efficiency differs fundamentally from a thermodynamic approach based on calculating the entropy budgets
under the assumption that the atmosphere works as a heat engine, e.g., \citep{pau02a,pau02b}, see also \citep{mgln10}.
The principal limitation of the entropy-budget
approach is that while the upper bounds on the amount of work that {\it could} be produced are clarified,
there is no indication regarding the degree to which such work {\it is} actually performed. In other words, the presence of an atmospheric
temperature gradient  is insufficient to  guarantee that mechanical work is produced.
In contrast, our estimate~(\ref{Pi}) is based on an explicit calculation of mechanical work
derived from a defined atmospheric pressure gradient. It is, to our knowledge,
the only available estimate of efficiency $\eta$ made from the basic physical parameters that characterize the atmosphere.

\subsection{Evaporation and condensation}

While condensation releases the potential energy of atmospheric water vapor,
evaporation, conversely, replenishes it. Here we briefly dwell on some
salient differences between evaporation and condensation to complete our picture regarding how the phase
transitions of water vapor generate pressure gradients.

Evaporation requires an input of energy to overcome the intermolecular forces of attraction
in the liquid water to free the water molecule to the gaseous phase, as well as to compress the air.
That is, work is performed against local atmospheric pressure to
make space for vapor molecules that are being added to the atmosphere via
evaporation. This work, associated with evaporation, is the source of potential
energy for the condensation-induced air circulation. Upon
condensation, two distinct forms of potential energy arise. One is
associated with the potential energy of raised liquid drops -- this potential energy dissipates to friction as
the drops fall. The second form of potential energy is
associated with the formation of a non-equilibrium pressure
gradient, as the removal of vapor from the gas phase creates
a pressure shortage of moist air aloft. This
pressure gradient
produces air movement. In the stationary case total
frictional dissipation in the resulting circulation is balanced
by the fraction of solar power spent on the work associated
with evaporation.

Evaporation is a surface-specific process. It is predominantly anchored to the Earth's surface.
In the stationary case, as long there is a supply of energy and the relative humidity is less than unity, evaporation
is adding water vapor to the atmospheric column without changing its temperature. The rate of evaporation is affected by turbulent mixing
and is usually related to the {\it horizontal} wind speed at the surface.
The global mean power of evaporation cannot exceed the power of solar radiation.

In contrast, condensation is a volume-specific, rather than an area-specific, process that affects the entire atmospheric column.
The primary cause of condensation is the cooling of air masses as the moist air ascends and its temperature drops.
Provided there is enough water vapor in the
ascending air, at a local and short-term scale condensation is not governed by solar power but by stored energy and can occur at an arbitrarily high rate
dictated by the {\it vertical} velocity of the ascending flow, see (\ref{src}).

Any circulation pattern includes areas of lower pressure where
air ascends, as well as higher pressure areas where it descends. Condensation rates are
non-uniform across these areas -- being greater in areas of ascent. Importantly, in such areas of ascent condensation involves
water vapor that is locally evaporated along with often substantial amounts of additional water vapor transported from elsewhere.
Therefore, the mean rate of condensation in the ascending region of any circulation pattern is {\it always} higher
than the local rate of evaporation. This inherent spatial non-uniformity of the condensation process determines horizontal pressure gradients.

Consider a large-scale stationary circulation where the regions of ascent and descent are of comparable size.
A relevant example would be the annually averaged circulation between the Amazon river basin (the area of ascent) and the region
of Atlantic ocean where the air returns from the Amazon to descend depleted of moisture. Assuming that the relative humidity at the surface,
horizontal wind speed and solar power are approximately the same in the two regions, mean evaporation rates should be roughly similar as well
(i.e., coincide at least in the order of magnitude). However, the condensation (and precipitation) rates in the two
regions will be consistently different. In accordance with the picture outlined above,
the average precipitation rate $P_a$ in the area of ascent should be approximately double the average value of regional evaporation rate $E_a$.
The pressure drop caused by condensation cannot be compensated by local evaporation to produce a net zero effect
on air pressure.
This is because in the region of ascent {\it both} the local water vapor evaporated from the forest canopy of the
Amazon forest at a rate $E_a \sim E_d$ as well as imported water vapor evaporated from the ocean
surface at a rate $E_d$ precipitate,
$P_a = E_d + E_a$. This is confirmed by observations: precipitation in the Amazon river basin
is approximately double the regional evaporation, $P_a \approx 2E_a$ \citep{mar04}. The difference between regional rates of precipitation and evaporation on land,
$R = P_a - E_a \sim E_a$, is equal to regional runoff. Note that in the region of descent the runoff thus defined is negative and
corresponds to the flux of water vapor that is exported away from the region with the air flow. Where runoff is positive, it represent the flux
of liquid water that leaves the region of ascent to the ocean.

The fact that the climatological
means of evaporation and precipitation are not commonly observed to be equal has been recognized in the literature \citep[e.g.,][]{wack03},
as has the fact that local mean precipitation values are consistently larger than those for evaporation  \citep[e.g.,][]{tr03}.

The inherent spatial non-uniformity of the condensation process  explains why it is condensation that principally determines the pressure gradients associated with
water vapor. So, while evaporation is adding vapor to the atmosphere and thus {\it increasing} local air pressure,
while condensation in contrast {\it decreases} it, the evaporation process is significantly more even and uniform spatially
than is condensation. Roughly speaking, in the considered example evaporation increases pressure near equally in the regions of ascent and descent,
while condensation decreases pressure only in the region of ascent. Moreover, as discussed above, the rate at which the air pressure is decreased
by condensation in the region of ascent is always higher than the rate at which local evaporation would increase air pressure.
The difference between the two rates is particularly marked in heavily precipitating systems like hurricanes, where precipitation
rates associated with strong updrafts can exceed local evaporation rates by more than an order of magnitude \citep[e.g.,][]{TF07}.

We have so far discussed the magnitude of pressure gradients that are produced and maintained by condensation in the regions
where the moist air ascends. This analysis is applicable to observed condensation processes that occur on different spatial scales, as we
illustrated on the example of Hadley Cell. We emphasize that to determine {\it where} the ascending air flow and condensation will
predominantly occur is a separate physical problem. For example, why the updrafts are located over the Amazon and
the downdrafts are located over the Atlantic ocean and not vice versa. Here regional evaporation patterns play a crucial role. In Section~4.1 we have shown
that constant relative humidity associated with surface evaporation, which ensures that $\pt N_v/\pt r = 0$, is necessary for the condensation to take
place. Using the definition of $\gamma$~(\ref{gamma}) equation~(\ref{gr2})
can be re-written as follows:
\beq
\label{contgr}
\frac{\pt \ln\gamma}{\pt r} = - \frac{w}{u}\frac{\pt \gamma}{\pt z}.
\eeq
\noindent
This equation shows that the decrease of $\gamma$ with height and, hence, condensation is only possible when $\gamma$
grows in the horizontal direction, $\pt \ln \gamma/\pt r > 0$. Indeed, surface pressure is lower in the region of ascent.
As the air moves towards the region of low pressure, it expands.
In the absence of evaporation, this expansion would make the water vapor contained
in the converging air unsaturated. Condensation at a given height would stop.

Evaporation adds water vapor to the moving air to keep water vapor saturated and sustain condensation.
The higher the rate of evaporation, the larger the ratio $w/u$ at a given $\pt \gamma/\pt z$ and,
hence, the larger the pressure gradient (\ref{gr2}) that can be maintained between the regions of ascent and descent.
A small, but persistent difference in mean evaporation $\Delta E < E$ between two adjacent regions, determines
the predominant direction of the air flow.
This explains the role of the high leaf area index of the natural forests in keeping evaporation
higher than evaporation from the open water surface of the ocean, for the forests to become the regions of low pressure
to draw moist air from the oceans and not vice versa \citep{mg07}.
On the other hand, where the surface is relatively homogeneous with respect to
evaporation (e.g., the oceanic surface), the spatial and temporal localization of condensation events can be
of random nature.

\conclusions[Discussion: Condensation dynamics versus differential heating in the generation of atmospheric circulation]

In Section 2 we argued that condensation cannot occur adiabatically at constant volume but is always accompanied
by a pressure drop in the local air volume where it occurs. We concluded that the statement that
"the pressure drop by adiabatic condensation is overcompensated by latent heat induced pressure rise of the air" \citep[p.~S12437]{poe09}
was not correct. In Section 3 we quantified the pressure change produced by condensation as dependent on altitude in a column in hydrostatic balance,
to show that in such a column the pressure drops upon condensation everywhere in the lower atmosphere up to several kilometers altitude, Fig.~1c.
The estimated pressure drop at the surface increases exponentially with growing temperature and amounts to over 20~hPa at 300~K, Fig.~1b.

In Section 4 we discussed the implications of the condensation-induced pressure drop for atmospheric dynamics. We calculated the
horizontal pressure gradients produced by condensation and the efficiency of the atmosphere as a dynamic machine driven by condensation.
Our aim throughout has been to persuade the reader that these implications are significant in numerical terms and deserve a serious discussion
and further analysis. We will now conclude our consideration by discussing the condensation-induced dynamics at the background
of differential heating, a physical mechanism that, in contrast to condensation, has received much attention as an air circulation driver.

Atmospheric circulation is only maintained if, in agreement with the energy conservation law, there is a pressure gradient to accelerate the air masses and
sustain the existing kinetic energy of air motion against dissipative losses. For centuries, starting from the works of Hadley and his predecessors,
the air pressure gradient has been qualitatively associated with the differential heating of the Earth's surface and the Archimedes force (buoyancy)
which makes the warm and light air rise, and the cold and heavy air sink. This idea can be illustrated by Fig.~1c, where the warmer
atmospheric column appears to have higher air pressure at some heights than the colder column. In the conventional paradigm, this is expected
to cause air divergence aloft away from the warmer column, which, in its turn, will cause a drop of air pressure at the surface and the resulting surface
flow from the cold to the warm areas. Despite the physics of this differential heating effect being straightforward in qualitative terms,
the quantitative problem of predicting observed wind velocities from the fundamental physical parameters has posed enduring difficulties.
Slightly more than a decade before the first significant efforts in computer climate modelling, \citet{bru44} as cited by \citet{lew98} wrote:

{\small
"It has been pointed out by many writers that it is impossible to derive a theory of the general circulation based on the known value of the solar
constant, the constitution of the atmosphere, and the distribution of land and sea . . . It is only possible to
begin by assuming the known temperature distribution, then deriving the corresponding pressure distribution,
and finally the corresponding wind circulation".}

Brunt's difficulty relates to the realization that
pressure differences associated with atmospheric temperature gradients cannot be fully transformed into kinetic
energy. Some energy is lost to thermal conductivity without generating mechanical work.
This fraction could not be easily estimated by theory in his era -- and thus it has remained to the present.
The development of computers and appearance of rich satellite observations have facilitated
empirical parameterizations to replicate circulation in numerical models.
However, while these models provide reasonable replication of the quantitative features of the general circulation
they do not constitute a quantitative physical proof that the the observed circulation is driven by
pressure gradients associated with differential heating.
As \citet[p.~48]{lor67} emphasized,
although "it is sometimes possible to evaluate the long-term influence of each process affecting some feature of the circulation by recourse to
the observational data", such knowledge "will not by itself constitute an explanation of the circulation, since it will not reveal why each process
assumes the value which it does".

In comparison to temperature-associated pressure difference, the pressure difference associated with water vapor removal from the gas phase
can develop over a surface of uniform temperature. In addition, this pressure difference is physically anchored to the lower atmosphere. Unlike the temperature-related
pressure difference, it does not demand the existence of some downward transport of the pressure gradient from the upper to the lower atmosphere
(i.e., the divergence aloft from the warmer to the colder column as discussed above) to explain the appearance of
low altitude pressure gradients and the generation of surface winds.

Furthermore, as the condensation-related pressure difference $\delta p_s$ is not associated with a temperature difference,
the potential energy stored in the pressure gradient can be nearly fully converted to the kinetic energy
of air masses in the lower atmosphere without losses to heat conductivity. This fundamental difference between the two mechanisms of pressure difference generation can be traced in hurricanes. Within the hurricane there is a marked pressure gradient
at the surface. This difference is quantitatively accountable by the condensation process \citep{mg09b}$^{\ref{arx}}$. In the meantime, the possible
temperature difference in the upper atmosphere that might have been caused by the difference in moist versus dry lapse rates between
the regions of ascent and descent is cancelled by the strong horizontal mixing \citep{mo06}. Above approximately 1.5~km
the atmosphere within and outside the hurricane is approximately isothermal in the horizontal direction \citep[Fig.~4]{mo06}.
Therefore, while the temperature-associated pressure difference above height $z_c$, Fig.~1c, is not realized in the atmosphere, the
condensation-associated pressure difference below height $z_c$ apparently is.

Some hints on the relative strengths of the circulation driven by differential heating compared to condensation-induced circulation can be
gained from evaluating wind velocities in those real processes that develop in the lower atmosphere without condensation.
These are represented by dry (precipitation-free) breezes (such as diurnal wind patterns driven by the differential heating of land versus sea surfaces) and dust devils.
While both demand very large temperature gradients (vertical or horizontal) to arise as compared to the global mean values, both circulation types
are of comparatively low intensity and have negligible significance to the global circulation. For example, dust devils do not involve precipitation
and are typically characterized by wind velocities of several meters per second \citep{sin73}. The other type of similarly compact rotating
vortexes -- tornadoes -- that are always accompanied by phase transitions of water -- develop wind velocities that are at least an order of magnitude
higher \citep{wur96}. More refined analyses of Hadley circulation \citep{hh80}
point towards the same conclusion: theoretically described Hadley cells driven by differential heating appear to be one order of magnitude weaker than the observed
circulation \citep{hh80,schn06}, see also \citep{ca08}. While the theoretical description of the general atmospheric circulation
remains unresolved, condensation-induced dynamics offers a possible solution (as shown in Section~4.1).

Our approach and theory have other significant implications. Some have been documented in previous papers, for example with regard
to the development of hurricanes \citep{mg09a,mg09b} and the significance of vegetation and terrestrial evaporation fluxes in determining large scale
continental weather patterns \citep{mgl06,mg07,doug09,mgl09}. Other implications are likely to be important in predicting the global and local nature of climate change -- a subject of
considerable concern and debate at the present time \citep{pi09,sc10}.

In summary, although the formation of air pressure gradients via condensation has not received
detailed fundamental consideration in climatological and meteorological sciences, here we have argued that this lack of attention has been undeserved.
Condensation-induced dynamics emerges as a new field of investigations that can significantly enrich our understanding of atmospheric
processes and climate change. We very much hope that our present account will provide a spur for
further investigations both theoretical and empirical into these important, but as yet imperfectly characterized, phenomena.

\begin{acknowledgements}
We thank D.~R.~Rosenfeld and H.~H.~G.~Savenije for disclosing their names as referees in the ACPD discussion of the work of \citet{mgl08} and
D.~R.~Rosenfeld for providing clarifications regarding the derivation of the estimate of condensation-related pressure change as given by
\citet[p.~S12436]{poe09}. We acknowledge helpful comments of K.~E.~Trenberth towards a greater clarity of the presentation of our approach.
The authors benefited greatly from an exciting discussion of condensation-related dynamics with J.~I.~Belanger, J.~A.~Curry, G.~M.~Lackmann,
A.~Seimon and R.~M.~Yablonsky.

\end{acknowledgements}

\end{document}